\documentclass[a4paper,12pt]{article}
\usepackage{graphicx}
\usepackage{latexsym}
\usepackage{float}
\usepackage{amsmath}
\usepackage{amsfonts}
\usepackage{psfrag}
\usepackage{dcolumn}

\def\blackboardrrm{\mathchoice
{\rm I\kern-0.21 em{R}}{\rm I\kern-0.21 em{R}}
{\rm I\kern-0.19 em{R}}{\rm I\kern-0.19 em{R}}}

\def\blackboardzrm{\mathchoice
{\rm Z\kern-0.32 em{Z}}{\rm Z\kern-0.32 em{Z}}
{\rm Z\kern-0.28 em{Z}}{\rm Z\kern-0.28 em{Z}}}
  
\newcommand{\Order}{\mathcal{O}} 
\newcommand{\Lag}{\mathcal{L}} 

\newcolumntype{d}[1]{D{,}{}{#1}}
\newcolumntype{e}[0]{D{.}{}{0}}
\newcolumntype{f}[1]{D{.}{\phantom{0}}{#1}}

\begin{document}

\title{
Numerical confirmation of analytic predictions for the finite volume mass gap of the XY-model
}

\author{
Janos Balog\\Research Institute for Particle and Nuclear Physics\\
1525 Budapest 114, Pf. 49, Hungary\\[1ex]
Francesco Knechtli, Tomasz Korzec and Ulli Wolff\\
Institut f\"ur Physik, Humboldt Universit\"at\\ 
Newtonstr. 15 \\ 
12489 Berlin, Germany
}
\date{}
\maketitle

\vspace*{1cm}

\begin{abstract}
Recent exact predictions for the massive scaling limit
of the two dimensional XY-model are based on the equivalence with the sine-Gordon 
theory and include detailed results on the finite size
behavior. The so-called step-scaling function of the mass gap
is simulated with very high precision and found consistent with
analytic results in the continuum limit.
To come to this conclusion, an also predicted form of a logarithmic
decay of lattice artifacts was essential to use for the extrapolation.
\end{abstract}

\vspace*{1cm}

\begin{flushright} HU-EP-03/54 \end{flushright}
\begin{flushright} SFB/CCP-03-28 \end{flushright}
\thispagestyle{empty}
\newpage


\section{Introduction} 
For two-dimensional quantum field theories there are exact
solutions available in quite a number of cases. The meaning of
the notion `exact' is however not quite universal. In many cases
it does not imply a straight-forward evaluation, and often not rigorously
provable conjectures are adopted at intermediate steps. To corroborate
such solutions  
it is very attractive to
compare --- whenever possible ---
high precision numerical simulations with exact predictions.
An interesting such opportunity arose in recent years from an
exact solution of the famous XY-model \cite{KosterlitzThouless73,Kosterlitz74} 
in its massive continuum limit
taken from the vortex-phase. This solution involves a conjectured
chain of equivalences that links it to the exactly solved sine-Gordon
field theory \cite{Zinn-Justin}. 
The predictions are particularly suitable
for a numerical check since they include information about
the XY-model in a finite volume and even about the asymptotic
lattice spacing dependence for the standard discretization of the model.
On the simulation side the XY-model profits from the practical absence of
critical slowing down for cluster algorithms \cite{Wolff89}, allowing for large
correlation lengths to be simulated, and from the availability of
reduced variance estimators for correlations \cite{Hasenbusch94_2}.

Interesting light is shed on a problem inherent in all numerical
lattice field theory computations, including QCD and asymptotically
free two-di\-men\-sional models: the need of a continuum extrapolation.
To carry it out, a theory based analytic form for the asymptotic
dependence on the lattice spacing $a$ is required. 
Usually a power behavior $a$ and/or $a^2$ from Symanzik
theory \cite{Symanzik83}, based on all-order perturbation theory,  is used.
In \cite{Hasenfratz2000,Hasenbusch2001} high precision simulations
of the $O(3)$-model revealed however that
the expected $a^2$ behavior does not prevail down to very small
lattice spacings, and at present there seems to be no theoretical
understanding for this. The XY-model is not asymptotically
free in the usual sense and the exact prediction for its $a$-dependence
differs from the Symanzik behavior, which is very interesting to
verify. A good understanding of the cut-off dependence is clearly
important also for QCD simulations, in particular now that they become
more and more precise and hence quantitatively useful for phenomenology.

\section{Theoretical predictions}
\subsection{The LWW-coupling}
   The two dimensional nonlinear $O(2)$ symmetric $\sigma$-model can be defined by
   the classical Lagrangian
   \begin{equation} \label{sigmaDefinition}
      \Lag = \frac{1}{2g^2}\partial_\mu S^a \partial_\mu S^a, \qquad S^aS^a =1, \qquad a = 1,2 \,.
   \end{equation}
   When space-time is discretized the XY-model is recovered, its standard action being
   \begin{equation} \label{CosineAction}
      S = -\beta \sum_{\langle k,l\rangle} \cos(\theta_k - \theta_l) \, .
   \end{equation}
   The summation is performed over all nearest neighbor pairs.  The model describes
   a set of spins with unit length arranged on the sites of a square lattice
   that are subject to a ferromagnetic interaction. The angle $\theta_n$ denotes
   the alignment of the spin at site $n$ relative to an arbitrary axis and 
   $\beta=1/g^2$
   is the inverse temperature or bare coupling.
   
   Let the spatial extent of the lattice be $L$. This introduces a natural external
   scale and renormalized
   couplings that run with $L$ can be defined. In~\cite{LWW91}
   the L\"uscher-Weisz-Wolff-coupling (LWW) was introduced
   \begin{equation} \label{LWWcoupling}
      \bar g^2(L) = 2M(L)L \, .
   \end{equation}
   Here $M(L)$ is the finite volume mass gap of the theory defined by the 
   transfer matrix in the sector
   of zero spatial momentum. 
   The running of the coupling
   is described by the $\beta$-function\footnote{The symbol $\beta$ is 
   traditionally used for both the 
   inverse temperature and the $\beta$-function which should not be confused 
   with each other.}
   \begin{equation} \label{betafun}
      \beta(\bar g^2) = -L \frac{\partial \bar g^2}{\partial L} \, .
   \end{equation}
   For finite scale changes
   the universal 
   step-scaling function $\sigma$ 
   has been defined, and it describes how the LWW coupling changes if the scale is
   dilated by a factor $s$
   \begin{equation}
      \sigma(s,\bar g^2(L)) = \bar g^2(sL) \, .
   \end{equation}
   The lattice version ($a$ is the lattice spacing) of the step scaling function $\Sigma(s,\bar g^2,a/L)$ 
   depends in addition on the resolution of the lattice.   
   
\subsection{DdV equation} 
   Starting from the light-cone lattice
   regularization \cite{DdV87,DdV89} of the sine-Gordon model and the 
   corresponding Bethe Ansatz solution, a complete description
   of the exact spectrum of the model in finite volume
   is provided by the Destri-deVega (DdV)
   non-linear integral equations. They were originally suggested~\cite{DdV92,DdV95}
   for the ground state of the sine-Gordon or the closely
   related massive Thirring model, but can be systematically
   generalized to describe excited states as well; the (common)
   even charge sector of both models~\cite{DdV97,Feverati98,Feverati99} and the (different)
   odd charge sector in both sine-Gordon and massive Thirring model \cite{Feverati98_2}.

   The ground state energy is given by
   \begin{equation}\label{DdV_E0}
      E_0(L) = -\frac{M}{\pi}\, {\rm Im} \int \limits_{-\infty}^{+\infty}\! \! dx \
               \sinh (x+i\eta)\, \ln \left[ 1+ e^{iZ_0(x+i\eta)}\right] \, .
   \end{equation}
   The mass
   $M$ without an argument denotes here
   the infinite volume mass gap $M \equiv M(L=\infty)$. The parameter $\eta$
   is needed to keep the branch of the logarithm well-defined. The result 
   is independent of $\eta$ as long as
   $0<\eta<\pi$.
   The so-called counting function $Z_0(\theta)$ solves the nonlinear integral equation
   \begin{equation} \label{DdV_Z0}
   \begin{split}
       Z_0(\theta) =& \ ML \sinh \theta \\ &- i\int \limits_{-\infty}^{+\infty} \! dx \
     \left[G(\theta - x - i\eta)-G(\theta + x + i\eta)\right]
      \ln \left[1+ e^{iZ_0(x+i\eta)}\right]
   \end{split}
   \end{equation}
   with a kernel given by
   \begin{equation}\label{DdV_kernelSG}
      G(\theta) = \frac{1}{2\pi}\int \limits_{-\infty}^{+\infty} \! \! dk \ e^{ik\theta}
      \ \frac{\sinh \frac{\pi k(p-1)}{2}}{2\, \cosh \frac{\pi k}{2}\sinh \frac{\pi k p}{2}} \, ,
   \end{equation}
   where $p=\beta^2_{\rm sG} /(8\pi-\beta_{\rm sG}^2)$ parameterizes the sine-Gordon coupling $\beta_{\rm sG}$.
      
   Similarly the energy of the first excited state is given by
   \begin{equation}\label{DdV_E1}
      E_1(L) = M-\frac{M}{\pi}\, {\rm Im} \int \limits_{-\infty}^{+\infty}\! \! dx \
               \sinh (x+i\eta)\, \ln \left[ 1- e^{iZ_1(x+i\eta)}\right]\, ,
   \end{equation}
   with $Z_1(\theta)$ being a solution of
   \begin{equation} \label{DdV_Z1}
   \begin{split}
       Z_1(\theta) =& \ ML \sinh \theta + \chi(\theta) \\ &- i\int \limits_{-\infty}^{+\infty} \! dx \
     \left[G(\theta - x - i\eta)-G(\theta + x + i\eta)\right]
      \ln \left[1- e^{iZ_1(x+i\eta)}\right] \, .
   \end{split}
   \end{equation}
   The source term $\chi(\theta)$ is of the form
   \begin{equation} \label{DdV_chi}
      \chi(\theta) = 2\pi \, \int \limits_{0}^{\theta} \!\! dz \, G(z)
   \end{equation}
   and  $\eta$ is arbitrary within the same range as before.
   
   The finite volume mass gap of the theory is $M(L) = E_1(L)-E_0(L)$.
   The XY-model is believed to lie in the same universality class as the
   sine-Gordon model with the sine-Gordon coupling $\beta_{\rm sG} = \sqrt{8\pi}$,
   or $p \to \infty$ \cite{Zinn-Justin}. In this case the kernel simplifies to
   \begin{equation} \label{DdV_kernelXY}
      G(\theta) = \frac{1}{2\pi} \int \limits_{-\infty}^{+\infty} \!\! dk \, \frac{e^{ik\theta}}{e^{\pi|k|}+1} \, .
   \end{equation}
   The Fourier integral  can be carried out
   \begin{equation} \label{DdV_kernelXY2}
      G(\theta) = \frac{1}{4\pi^2}\left(-\Psi(\frac{\pi-i\theta}{2\pi})- \Psi(\frac{\pi+i\theta}{2\pi})
                                +\Psi(1-\frac{i\theta}{2\pi}) + \Psi(1+\frac{i\theta}{2\pi}) \right)
   \end{equation}
   in terms of the
   digamma function\footnote{The digamma function $\Psi(z)$ is the logarithmic derivative of the gamma function,
   given by $\Psi(z) =\Gamma'(z)/\Gamma(z)$.} $\Psi(z)$. 
   Also the source term can be calculated exactly in the limit $p \to \infty$,
   \begin{equation} \label{DdV_chiXY}
   \begin{split}
      \chi(x+i\eta) =& -i\, \ln\left[ \Gamma
\left( \frac{1}{2}-\frac{ix-\eta}{2\pi}\right) \right]
                      +i\, \ln\left[ \Gamma
\left( \frac{1}{2}+\frac{ix-\eta}{2\pi}\right) \right] \\
                     & +i\, \ln\left[ \Gamma\left( 1-\frac{ix-\eta}{2\pi}\right) \right]
                      -i\, \ln\left[ \Gamma\left( 1+\frac{ix-\eta}{2\pi}\right) \right] \, .
   \end{split}
   \end{equation}

\subsection{Asymptotic behavior}
   For large volumes ($L\to\infty$) and also for small volumes ($L\to0$)
   asymptotic formulae exist which can be used to calculate the finite volume 
   mass gap in this model independently of the DdV equation.

   The $\beta^2_{\rm sG}\to8\pi$ sine-Gordon model is equivalent
   to the $SU(2)$ symmetric chiral Gross-Neveu model and is thus
   asymptotically free. In principle it is straightforward to
   do perturbation theory for the XY-model using the formalism
   of Amit et al. \cite{Amit79}. The final result, as usual in an
   asymptotically free model, is a power series in a running coupling.
   Technically however, the calculations are more involved than usual,
   since the perturbative contributions cannot be written in terms of
   usual Feynman integrals.

   The 2-loop perturbative result for the LWW coupling is
   \begin{equation} \label{shortDistance}
      \bar g^2 = \frac{\pi}{2}(1+\gamma+u_2 \gamma^2) + \Order(\gamma^3) \, ,
   \end{equation}
   with the running coupling $\gamma$ solving
   \begin{equation}
      \frac{1}{2\gamma}-\frac{1}{2} \ln (2\gamma) = \ln \left( \frac{1}{ML}\right).
   \end{equation}
   Along the lines described in \cite{Amit79} we obtain
   \begin{equation}
   u_2=0.20(1) ,
   \end{equation}
   where due to the complexity of the computation the 2-loop coefficient has to be obtained
   numerically in the end. 

   On the other hand, L\"uscher has found a universal
   formula which describes the leading large volume corrections
   to the mass of the lightest particle on a spatial torus in terms of scattering
   amplitudes of the theory \cite{Luescher86}.
   The formula exists in any dimension.
   It is particularly useful in 
   two dimensional integrable models where the scattering amplitudes are
   known exactly. In our case it gives
   \begin{equation} \label{LueschersFormula}
   \begin{split}
      \bar g^2 =& 2ML + \frac{2ML}{\pi} \int \limits_{-\infty}^{+\infty} \! \! d\theta \ \cosh \theta \, e^{-ML \cosh \theta}
               \left[ 1+ e^{i\chi(\theta+\frac{i\pi}{2})}\right] \\ &+ \Order (e^{-2ML}).
   \end{split}
   \end{equation}

\section{Numerical results}
\subsection{Numerical solution of the DdV equation}   
\begin{figure} [h]
     \centering
     \includegraphics [width = \textwidth] {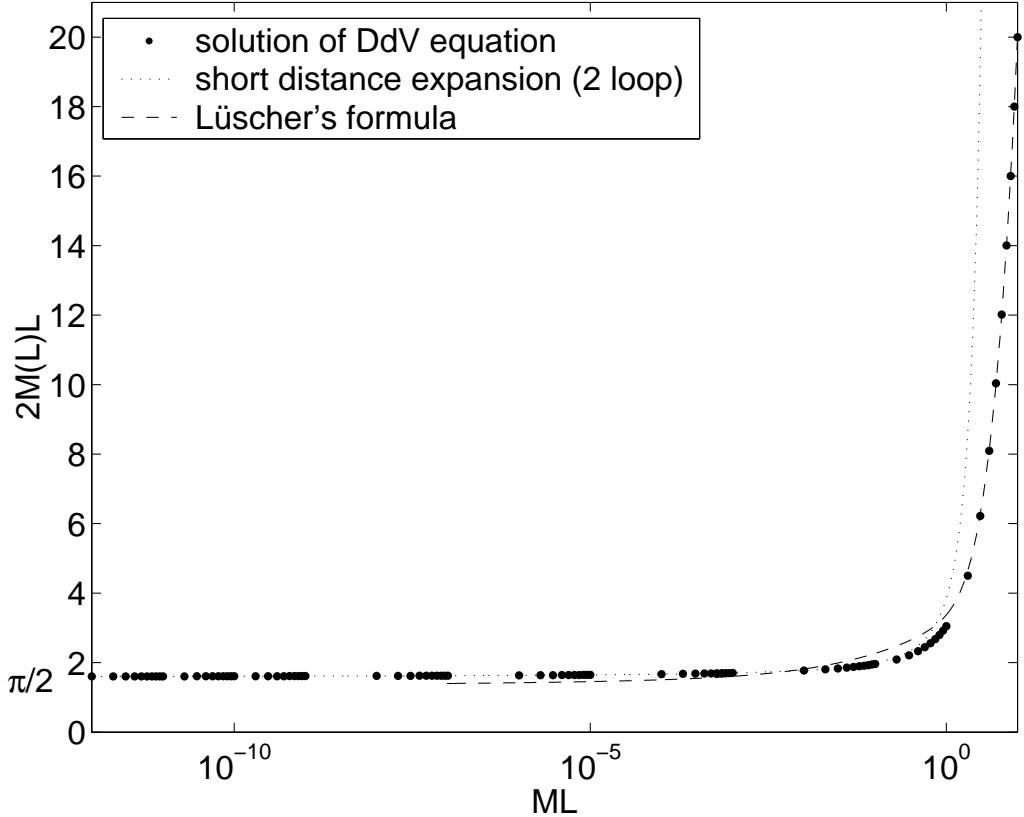}
     \caption{LWW coupling obtained from solutions of the DdV equation.}
     \label{gbarddvfig}
   \end{figure}

   To calculate the finite volume mass gap $M(L)$ it is necessary to
   evaluate the nonlinear integral equations (\ref{DdV_Z0}) and (\ref{DdV_Z1}) numerically.
   The idea is to solve them iteratively \cite{Ravanini2001}. In a first step the 
   integral term on the right hand side is neglected, in every consecutive step the 
   previous approximation is inserted 
   into the integral. The iteration is aborted when the desired precision is reached. 
   Further details of this 
   procedure are described in \cite{korzec2003}. 
   Figure \ref{gbarddvfig} shows some values of the LWW-coupling 
   as a function of the the size in units of the infinite volume mass gap
   obtained from the
   numerical solutions of  the DdV equations and compares them to the asymptotic 
   formulas~(\ref{shortDistance}) and~(\ref{LueschersFormula}).

   To compute the step-scaling function $\sigma(2,u)$ first a value $ML$
   has to be found such that after solving the DdV equation for it
   $2M(L)L=u$ holds accurately enough.
   Then $\sigma$ is given by solving again with the doubled value $2ML$.

\subsection{Monte Carlo study of the model}
   The XY-model can be simulated very effectively with cluster algorithms~\cite{Wolff89,Wolff88,Niedermayer97}.
   The finite volume mass gap may be obtained from the decay of the 
   correlation function of spatially averaged spin fields (see \cite{LWW91} for details).
   For our simulations we use a single cluster algorithm and an improved estimator for
   the correlation function \cite{Hasenbusch94_2}.

   To calculate the step scaling function we perform calculations on lattices of
   different resolutions $a/L$. For each lattice a bare coupling $\beta$ is
   determined that leads to the desired value $u$ of the LWW-coupling. A simulation at the
   same bare coupling but with lattice size $2L/a$ yields a point of the 
   lattice step-scaling function $\Sigma(2,u,a/L)$. To obtain the continuum step-scaling function it is
   necessary to perform a continuum extrapolation. The unusual form of the lattice artifacts 
   in the XY-model was predicted in \cite{Balog2000} and in the case of the step-scaling
   function reads
   \begin{equation} \label{latart}
       \Sigma(2, \bar g^2, a/L ) = 
       \sigma(2,\bar g^2) + \frac{c}{(\ln \xi + U)^2} + \Order(\ln^{-4}\xi) \, ,
   \end{equation}
   where $U$ is a non-universal constant ($U=1.3(1)$ for the standard 
   action \cite{Balog2002}) and
   $c$ can be calculated for each $\bar g^2$ 
   (see appendix \ref{latartAppendix}). The 
   lattice resolution is expressed in terms of the infinite volume correlation length
   $\xi = \frac{1}{Ma}=\frac{L/a}{ML}$, which can be measured directly if it is not too large.   
   When the bare coupling $\beta$ approaches $\beta_c$, the Kosterlitz Thouless (KT) 
   theory~\cite{KosterlitzThouless73,Kosterlitz74} predicts the correlation length to 
   diverge according to
   \begin{equation} \label{xiKT}
	\xi = A \exp \left( C \left| \frac{\beta_c-\beta}{\beta_c} \right|^{-\frac{1}{2}}\right) \, .
   \end{equation}
   The non-universal constants are known with good precision for the standard action  \cite{Hasenbusch97},
   \begin{eqnarray}
      A       &=& 0.233 \pm 0.003 \\
      C       &=& 1.776 \pm 0.004 \\
      \beta_c &=& 1.1199 \pm 0.0001 \, .
   \end{eqnarray}
   This result was derived in an approximative renormalization group study 
   and is strictly valid only for
   $\beta \to \beta_c$. At $\beta \sim 1.05$ we see a numerical
   matching with KT, so we use eq. (\ref{xiKT})
   to determine the correlation lengths yet closer to criticality.
   We emphasize that this result is only used for the lattice artifacts
   and that errors in $\xi$ are suppressed by the logarithm in (\ref{latart}). 

\begin{figure} [h]
	 \centering
	 \psfrag{tag1}{$(1.3+\ln \xi)^{-2}$}
	 \psfrag{tag2}{$\Sigma(2,\bar g^2 = 1.7865, a/L)$}
	 \psfrag{tag3}{MC-data}
	 \psfrag{tag4}{solution of DdV equation}
	 \psfrag{tag5}{linear fit}
	 \includegraphics [width = \textwidth] {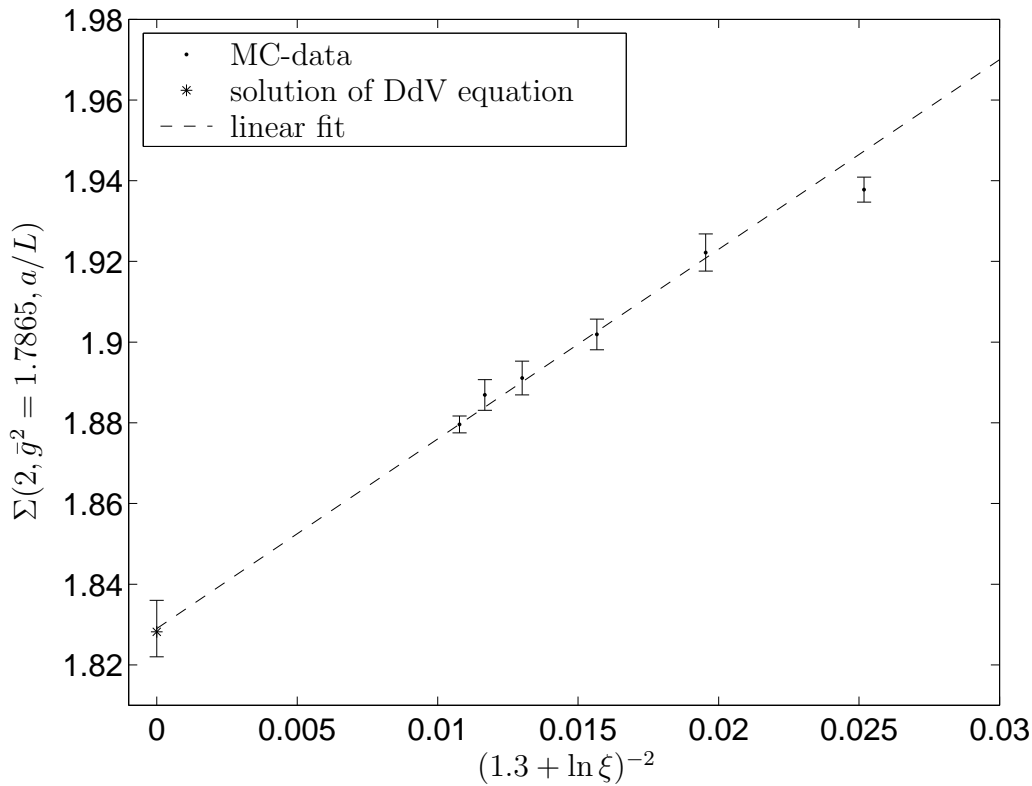}
	 \caption{Comparison of MC-data with a numerical solution of the DdV equation at $\bar g^2 = 1.7865$.
        	  The spatial extents $L/a$ of the lattices were $10,\ 20,\ 40,\ 80,\ 120$ and $160$.
        	  The smallest lattice was discarded in the fit.}
	 \label{MCvsDdVFig1}
   \end{figure}

\begin{figure} [h]
	\centering
	\psfrag{tag1}{$a/L$}
	\psfrag{tag2}{$\Sigma(2,\bar g^2 = 1.7865, a/L)$}
	\psfrag{tag3}{MC-data}
	\psfrag{tag4}{solution of DdV equation}
	\psfrag{tag5}{polynomial fit}
	\includegraphics [width = \textwidth] {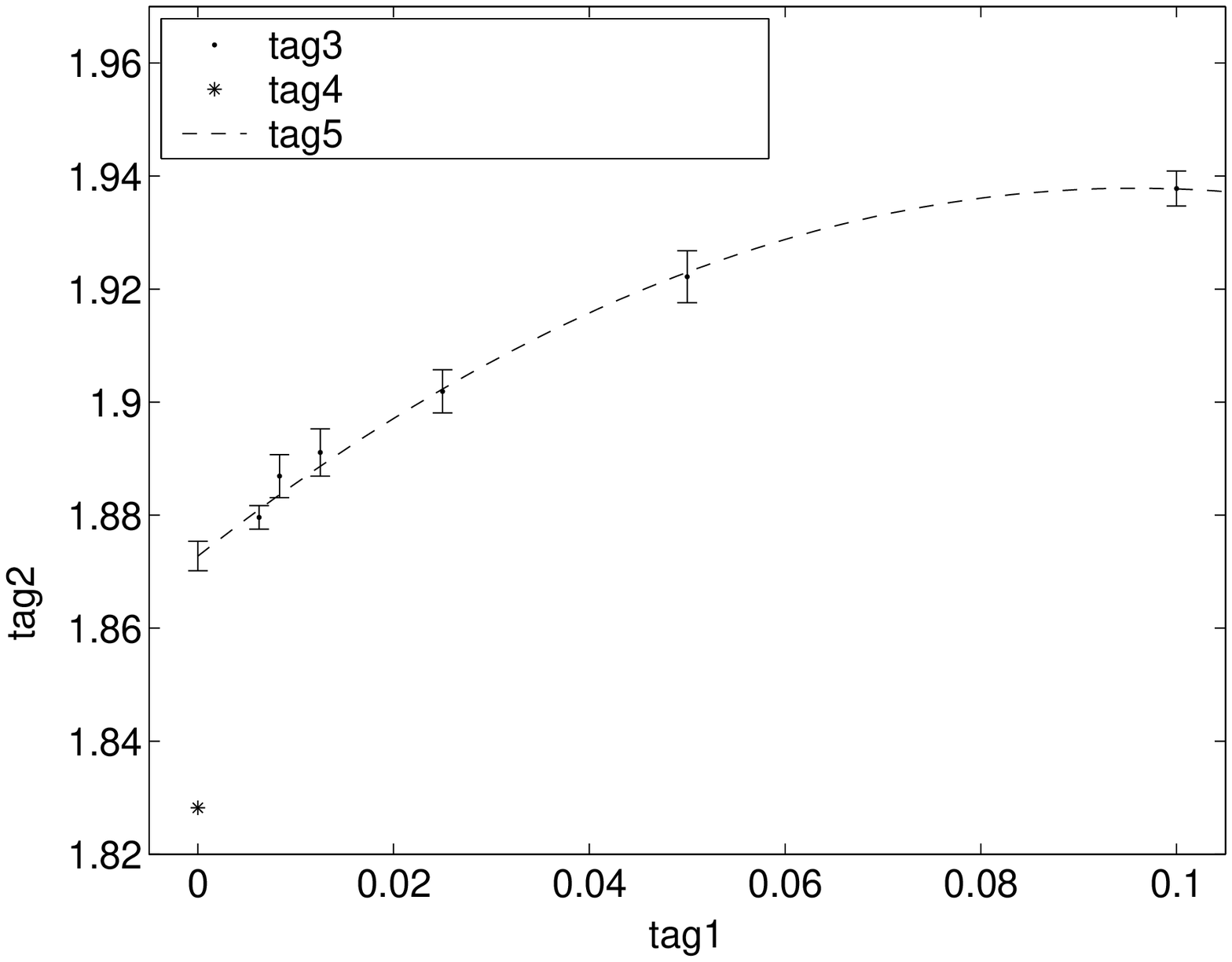}
	\caption{Comparison of MC-data with a numerical solution of the DdV equation at $\bar g^2 = 1.7865$.
        	 The predicted continuum value could not have been confirmed
        	 by an extrapolation with a second degree polynomial (including a linear term),
		 as used in this plot.}
	\label{MCvsDdVFig2}
   \end{figure}

   Fitting the Monte Carlo estimates for the lattice step-scaling function to (\ref{latart}) yields
   a point of the continuum step scaling function and the constant $c$. Both values can be compared
   to predicted values, which is done for several values of the LWW-coupling in table~\ref{stepscaleResults1}.
   The small discrepancy between $c_{\rm th}$ and $c_{\rm MC}$ can presumably be attributed to subleading
   cutoff effects in the Monte Carlo data.
   Figure \ref{MCvsDdVFig1} illustrates the extrapolation for one of the points, the corresponding figures 
   for the other points look qualitatively the same. Figure \ref{MCvsDdVFig2}, which has an equally good
   $\chi^2$-value, demonstrates how important the
   knowledge of the lattice artifacts is. The data needed to extract these results 
   are collected in table \ref{MCtable}
   of appendix \ref{tableAppendix}.

   \begin{table}[h]
	 \begin{center}
	 \begin{tabular}{|c| c c | c c | c|}
	 \hline
	 $ \bar g^2 $&$ \sigma_{\rm DdV}(2,\bar g^2) $&$ \sigma_{\rm MC}(2,\bar g^2)$&$  c_{\rm th} $&$  c_{\rm MC} $&$ \chi^2/_{\rm dof} $\\
	 \hline
	 $ 3.0038   $&$  4.3895  $&$ 4.40\phantom{0} \pm 0.02\phantom{0}$&$ 2.6176 \pm 0.0002 $&$  2.4 \pm 0.6 $&$   2.51 / 3    $\\
	 $ 1.7865   $&$  1.8282  $&$ 1.829 \pm 0.007                    $&$ 5.30\phantom{00} \pm 0.01\phantom{00} $&$  4.8 \pm 0.5 $&$   0.73 / 3    $\\
	 $ 1.6464   $&$  1.6515  $&$ 1.657 \pm 0.003                    $&$ 5.4\phantom{000} \pm 0.2\phantom{000} $&$  4.3 \pm 0.3 $&$   0.35 / 3    $\\
	 $ 1.6020   $&$  1.6029  $&$ 1.608 \pm 0.004                    $&$ 5.5\phantom{000} \pm 1.5\phantom{000} $&$  4.4 \pm 0.5 $&$   0.90 / 3    $\\  
	 \hline
	 \end{tabular}
	 \caption{At different values of the LWW-coupling theoretical predictions (DdV) for the step-scaling
	 function are compared with numerical results (MC). Also the constant $c$ 
	 as predicted by
	 theory (th) is compared to the numerical value.} \label{stepscaleResults1}
	 \end{center}
   \end{table}   

\section{Conclusions}  
   In this work we have found a remarkable consistency between analytic and
   numerical results. For several values of the LWW coupling we found perfect
   agreement between numerical continuum-extrapolated values of the step scaling function
   and those following from solving the DdV equation. This agreement is only
   obtained by employing for the extrapolation a logarithmic dependence on the
   lattice spacing that follows from the connection between the sine-Gordon
   and the XY-model together with the KT-behavior of the correlation
   length close to the critical point.

   This overall consistency furnishes strong evidence
   that all the considerations and assumptions
   leading from the six-vertex model
   via Bethe Ansatz to a set of integral equations that can be used to calculate the
   energy spectrum of the sine-Gordon model are correct.
   Simultaneously it also corroborates the assertion that the massive
   continuum limit  of the lattice XY-model can be described by the
   $\beta^2_{\rm sG}\to8\pi$ limit of the continuum sine-Gordon model.
   In Ref. \cite{Balog2002} the latter was investigated by the Form Factor
   Bootstrap construction. The renormalized 4-point coupling $g_{\rm R}$
   and 2-point correlation functions were compared to their lattice counterparts.
   Similarly to our findings here, after (and only after) the logarithmic
   lattice artifacts were taken into account, non-trivial agreement between
   lattice data and Form Factor results was found.

   Indirectly also the KT-scenario is confirmed once more.
   Calculations leading to this scenario are more or less approximate and there has been a
   controversy whether they can be trusted or not for a long time \cite{Luther77}. 
   This result
   demonstrates in an impressive way, how important it is to have 
   theory-based  information on the functional form
   of lattice artifacts in order to extract the right continuum values from 
   numerical data. 
   Although the accessible range of lattices is much bigger 
   and the precision higher than it would be in four 
   dimensional theories, an extrapolation
   to the continuum could lead to uncontrolled systematic errors 
   for precise data,
   if the right form of the lattice artifacts were not known.
   
{\bf Acknowledgement.} We would like to thank Burkhard Bunk for discussions.
This investigation was supported in part by the
Hungarian National Science Fund OTKA (under T034299
and T043159) and by the German science foundation DFG in Graduiertenkolleg GK271
and SFB/TR9.

\appendix
\section{Lattice artifacts of the step-scaling function} \label{latartAppendix}
   The LWW-coupling $\bar g^2 = 2M(L)L$ on a lattice is affected by lattice artifacts. 
   In contrast to the simulations we now consider families of lattice calculations 
   at fixed physical volume in units of the
   infinite volume mass gap, that is fixed $ML$. Then each chosen resolution $a/L$
   in principle fixes a value for $\beta$ and hence also $\bar g^2$.
   In this sense we find $\bar g^2 = f(ML,\xi)$, where $\xi=(Ma)^{-1}$ is used.
   The leading  $\xi$-dependence
   is predicted by theory~\cite{Balog2000} to be
   \begin{equation} \label{gbarArtifacts}
      f(ML,\xi) = f_0(ML) - \frac{f_1(ML)\pi^2}{2(\ln \xi +U)^2} + \Order([\ln \xi]^{-4}), 
      \quad f_0(ML) = f(ML,\infty)
   \end{equation}
   where $U$ is an action dependent constant ($U \approx 1.3$ for the standard action)
   and $f_1(ML)$ a coefficient. In solving the DdV equation (see figure \ref{gbarddvfig}) we evaluated
   $f_0(ML)$.   
   In \cite{Balog2000} it was shown that the same
   function $f_1$ also characterizes the approach of the
   the sine-Gordon model to its XY-model limit. 
   Here one may also define an LWW coupling $\bar g^2_{\rm sG}$
   and study its dependence on $p$ for fixed $ML$
   to derive
   \begin{equation} \label{f1fromSG}
      \bar g^2_{\rm sG} = \tilde f (ML,p) = f_0(ML)  + 
                          \frac{2f_1(ML)}{p^2} + \Order(p^{-4})
   \end{equation}
   for large $p$. This relation is entirely
   in the continuum and it hence becomes clear that $f_1$ is universal.

   Numerically we avoid reference to infinite volume 
   quantities (except here for the
   artifacts) and determine
   the step-scaling function $\sigma(2,\bar g^2)$ at a certain fixed $\bar g^2=u$. 
   The relation to the present language is
   \begin{equation}
   \sigma(2,u) = f_0[2 f_0^{-1}(u)],
   \end{equation}
   where $f_0^{-1}$ is the inverse function of $f_0$.
   At finite $\xi$ a computable slightly different value 
   $(M+\delta M)L$ is associated
   with the target $u$-value. Then the doubled value 
   is mapped to $\Sigma$ instead of $\sigma$.
   Collecting now all these corrections to leading order
   we obtain
   \begin{eqnarray}
   \Sigma(2,u,a/L) - \sigma(2,u) &=& 
     \left(
     \frac{2f_0'(2ML)}{f_0'(ML)} f_1(ML) - f_1(2ML)
     \right) \frac{\pi^2}{2(\ln \xi +U)^2} \nonumber \\
    &=& c(ML) \; \frac{1}{(\ln \xi +U)^2}
   \label{latartssf}
   \end{eqnarray}
   which defines $c(ML)$.

   Further corrections to (\ref{latartssf}) are of order $\Order([\ln \xi]^{-4})$.
   The required derivatives of $f_0$ in the continuum
   can be obtained numerically from solutions of the DdV equation. Also
   the coefficient $f_1$ may be gained from a numerical solution of the sine-Gordon version 
   of the DdV equations. 
   Instead of using the XY-model specific integral
   kernel (\ref{DdV_kernelXY2}) the version (\ref{DdV_kernelSG}) with finite $p$ is taken. 
   All data needed to 
   calculate $c$ is collected in table \ref{DdVtable}. \newpage

\section{Tables} \label{tableAppendix}
   In the following two tables some of the data is collected that is needed to extract the results. 
   An estimate of the derivative $\frac{\partial \Sigma}{\partial \bar g^2}$ is used to 
   propagate into the total error of $\Sigma$ the statistical error of $\bar g^2$ and to
   correct the value of $\Sigma$ for the small difference between $u$ and $\bar g^2$.
   In most cases this correction amounts to a small
   shift compared with the statistical errors. The results in table \ref{DdVtable} were obtained
   twice independently.
   
   \begin{table}[H]  
      \begin{center}        
      \begin{tabular}{|e d{5} d{9} d{9} d{4} d{8}|}  
      \hline
       L/a. & ,\phantom{0}\beta  & ,\phantom{0}\bar g^2(L)    & ,\bar g^2(2L)    & ,\phantom{0}u      & \Sigma(,2,u,a/L) \\        
      \hline            
      \hline
        20.  & 0.,9649  & 3.,0022(45)      & 4.,5144(74)       & 3.,0038 & 4.,519(13)  \\
        40.  & 1.,0000  & 3.,0036(41)      & 4.,4824(75)       & 3.,0038 & 4.,483(12)  \\
        80.  & 1.,0241  & 3.,0146(39)      & 4.,5055(76)       & 3.,0038 & 4.,480(12)  \\
       120.  & 1.,0359  & 2.,9919(39)      & 4.,4232(64)       & 3.,0038 & 4.,451(11)  \\
       160.  & 1.,0423  & 3.,0069(38)      & 4.,4742(64)       & 3.,0038 & 4.,466(11)  \\
      \hline 
        10   & 1.,0358  & 1.,7866(16)      & 1.,9379(19)       & 1.,7865 & 1.,9378(31)  \\
        20   & 1.,0541  & 1.,7837(16)      & 1.,9161(19)       & 1.,7865 & 1.,9222(40)  \\
        40   & 1.,0668  & 1.,7860(16)      & 1.,9010(19)       & 1.,7865 & 1.,9019(35)  \\
        80   & 1.,0756  & 1.,7896(16)      & 1.,8973(19)       & 1.,7865 & 1.,8911(37)  \\
       120   & 1.,0800  & 1.,7882(17)      & 1.,8898(19)       & 1.,7865 & 1.,8869(38)  \\
       160   & 1.,0830  & 1.,7850(16)      & 1.,8778(13)       & 1.,7865 & 1.,8796(21)  \\ 
      \hline
        20   & 1.,0758  & 1.,64585(73)     & 1.,71173(79)      & 1.,6464 & 1.,7125(12) \\
        40   & 1.,0844  & 1.,64632(75)     & 1.,70116(83)      & 1.,6464 & 1.,7013(16) \\
        80   & 1.,09076 & 1.,64646(73)     & 1.,69317(81)      & 1.,6464 & 1.,6931(13) \\
       120   & 1.,0937  & 1.,64636(74)     & 1.,68885(79)      & 1.,6464 & 1.,6889(13) \\
       160   & 1.,09551 & 1.,64645(77)     & 1.,68774(80)      & 1.,6464 & 1.,6877(13) \\  
      \hline              
        10   & 1.,0720  & 1.,5971(12)      &  1.,6721(66)      & 1.,602  &  1.,6801(69) \\
        20   & 1.,0825  & 1.,6033(12)      &  1.,6574(13)      & 1.,602  &  1.,6556(21) \\
        40   & 1.,0903  & 1.,6032(12)      &  1.,6463(14)      & 1.,602  &  1.,6449(20) \\
        80   & 1.,0960  & 1.,6020(11)      &  1.,6397(13)      & 1.,602  &  1.,6397(20) \\
       120   & 1.,0984  & 1.,6034(13)      &  1.,6365(14)      & 1.,602  &  1.,6349(20) \\
       160   & 1.,1000  & 1.,6028(13)      &  1.,6330(13)      & 1.,602  &  1.,6321(20) \\ 
      \hline
      \end{tabular}  
      \caption{Simulation parameters and results. In the last column all sources of errors have been 
       combined.} \label{MCtable}  
      \end{center}  
   \end{table}  
   \newpage
   
   \begin{table}[H]
      \begin{center}      
      \begin{tabular}{|d{11} c c d{9} d{11}|}
      \hline
       ,ML                    &$ p      $&$ \bar g^2_{\rm sG} $& ,f_1  & ,\frac{\partial f_0}{\partial ML}\\
      \hline \hline
       2.,164\times 10^{-12}  &$ 100    $&$ 1.60461349        $&       & \\
                              &$ 200    $&$ 1.60266193        $&       & \\
                              &$ 300    $&$ 1.60229497        $&       & \\
                              &$ \infty $&$ 1.60200011        $& 13.,1(2)   & 5.,616(4) \times 10^8  \\ 
      \hline
       4.,328\times 10^{-12}  &$ 100    $&$ 1.60541070        $&       & \\
                              &$ 200    $&$ 1.60350926        $&       & \\
                              &$ 300    $&$ 1.60315200        $&       & \\
                              &$ \infty $&$ 1.60286501        $& 12.,7(2)   & 2.,962(4) \times 10^8 \\
      \hline      
       9.,11895\times 10^{-6} &$ 100    $&$ 1.64750389        $&       & \\
                              &$  200   $&$ 1.64667661        $&       & \\
                              &$  300   $&$ 1.64652300        $&       & \\
                              &$ \infty $&$ 1.64640000        $&  5.,52(2)  & 7.,5826(6) \times 10^2 \\
      \hline
       1.,82379\times 10^{-5} &$ 100    $&$ 1.65254276        $&       & \\
                              &$ 200    $&$ 1.65176684        $&       & \\
                              &$ 300    $&$ 1.65162280        $&       & \\
                              &$ \infty $&$ 1.65150749        $&  5.,18(2)  & 4.,3054(4) \times 10^2 \\             
      \hline      
       0.,014024              &$ 100    $&$ 1.78689383        $&       & \\
                              &$ 200    $&$ 1.78659943        $&       & \\
                              &$ 300    $&$ 1.78654489        $&       & \\
                              &$ \infty $&$ 1.78650122        $&  1.,963(1)    & 3.,6430(4)   \\ 
      \hline
       0.,028048              &$ 100    $&$ 1.82853577        $&       & \\
                              &$ 200    $&$ 1.82828893        $&       & \\
                              &$ 300    $&$ 1.82824321        $&       & \\
                              &$ \infty $&$ 1.82820661        $&  1.,646(1)    & 2.,5230(2)   \\
      \hline
       0.,9652                &$ 100    $&$ 3.00389085        $&       & \\
                              &$ 200    $&$ 3.00385329        $&       & \\
                              &$ 300    $&$ 3.00384633        $&       & \\
                              &$ \infty $&$ 3.00384076        $&  0.,25041(1)  & 1.,27592(2)  \\
      \hline
       1.,9304                &$ 100    $&$ 4.38947652        $&       & \\
                              &$ 200    $&$ 4.38946271        $&       & \\
                              &$ 300    $&$ 4.38946015        $&       & \\
                              &$ \infty $&$ 4.38945811        $&  0.,092061(3) & 1.,58593(6)  \\
      \hline
      \end{tabular}
      \caption{Solutions of the DdV equation needed to calculate $c_{\rm th}$. Numerical values of $\bar g^2_{\rm sG}$
               are uncertain by one in the last digit.} \label{DdVtable}
      \end{center}
   \end{table}

\newpage
\bibliographystyle{h-elsevier}
\bibliography{references}
\end{document}